\documentclass[aip,cha,graphicx,amsmath,amssymb,reprint]{revtex4-2}

\usepackage{bm}
\usepackage{graphicx}
\usepackage{xcolor}
\graphicspath{{figures/}}

\begin{document}
\raggedbottom

\title{Axis-switching oscillations of elliptic scroll rings in excitable media}

\author{Oliver Steinbock}
\email{osteinbock@fsu.edu}
\affiliation{Department of Chemistry and Biochemistry, Florida State University, Tallahassee, Florida 32306-4390, USA}

\date{\today}

\begin{abstract}
Scroll waves in three-dimensional excitable media rotate around a filament that moves according
to its local curvature. Using simulations of the Barkley model, we show that initially planar,
elliptic scroll rings perform the axis-switching oscillation known from elliptic vortex rings in
fluids: while the ring drifts along its axis, it buckles into a saddle, becomes planar again with
interchanged axes, and repeats this cycle. The oscillation occurs for vanishing, positive, and
negative filament tension. As predicted by a kinematic filament equation, its period scales
approximately with the square of the ring radius and inversely with the drift coefficient. For our main model
parameters, however, the oscillation is about 1.45 times slower than the linear prediction based on
the drift of circular rings of the same size. A curvature-dependent drift coefficient accounts for a factor of
about 1.25, while the remaining slowdown results mainly from a weaker coupling of out-of-plane to
in-plane deformations, which a fourth-order rigidity term does not describe. In contrast, the
Panfilov--Rudenko model and the Atri model of Ca$^{2+}$ waves follow the linear prediction within
about 6\%, showing that the slowdown depends on the kinetics.
\end{abstract}

\maketitle

\begin{quotation}
Rotating spiral waves of chemical or electrical activity occur in systems as different as the
Belousov--Zhabotinsky reaction and the human heart, where they are associated with dangerous
arrhythmias. In three-dimensional media, a spiral extends into a scroll wave that rotates
around a curve called the filament. When the filament forms a closed loop, it travels
along its axis, much like a vortex ring in a fluid. For these fluid rings, J.~J. Thomson showed in 1883 that
a slight elliptic deformation causes an oscillation in which the long and short axes repeatedly exchange
places. Using
numerical simulations of excitable reaction--diffusion models, we show here that elliptic scroll rings
undergo the same oscillation. Its period scales with ring size approximately as predicted by the equation of
motion that also describes vortex filaments in fluids, but in one of the models it is markedly
longer than this equation predicts. Elliptic scroll rings therefore expose effects beyond the
standard description of filament motion.
\end{quotation}

\section{Introduction}
\label{sec:intro}

Excitable media relay a signal by local excitation followed by a period of recovery. The
best-known examples are nerve fibers and heart tissue, and similar excitation waves occur in the
Belousov--Zhabotinsky (BZ) reaction,\cite{Zaikin1970,Winfree1972,Winfree1973} as calcium waves
inside living cells,\cite{Lechleiter1991} and in populations of signaling slime mold
cells.\cite{Steinbock1993}
In two-dimensional excitable media, a wave with a free end typically curls into a rotating spiral. The
tip of the spiral is a phase singularity as all phases of the local excitation cycle meet at this
point. In three dimensions, these singular points form a continuous line, the filament, around
which the pattern rotates as a scroll wave.\cite{Winfree1973} The filament organizes the surrounding
three-dimensional wave field, and its motion is hence central to understanding how scroll waves
drift, collide, and break up.\cite{BiktashevReview2014} In the heart, such rotating waves are
associated with tachycardia and fibrillation.\cite{Davidenko1992,Gray1995}

\begin{figure*}
\includegraphics[width=\textwidth]{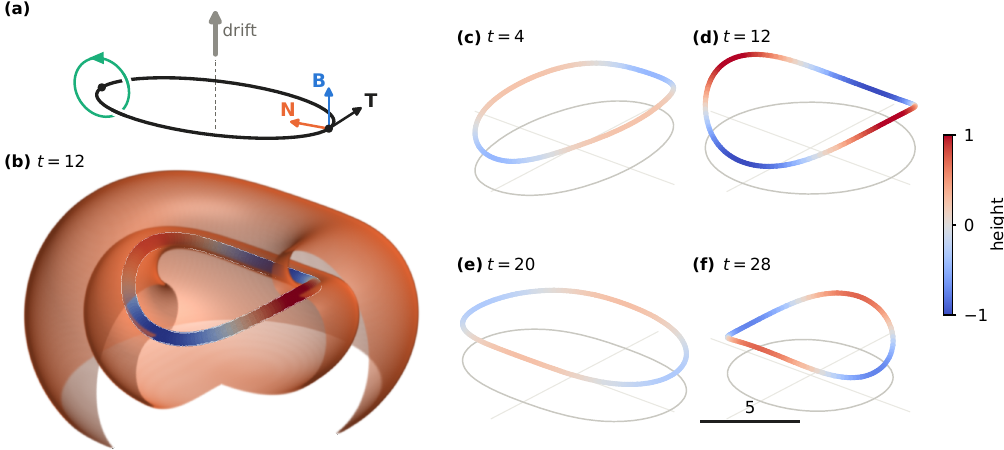}
\caption{\label{fig:cycle} Axis switching of an elliptic scroll ring. (a) Planar elliptic
filament (black) with its tangent $\bm{T}$, normal $\bm{N}$, and binormal $\bm{B}$ at the end of
the long axis, where the curvature is largest. The scroll wave rotates around the filament in the
plane spanned by $\bm{N}$ and $\bm{B}$, as indicated at the opposite end of the long axis (green
arrow). The binormal motion $\beta K\bm{B}$, averaged around the ring, makes the ring drift along
its axis (gray arrow). The filament is oriented by the right-hand rule (fingers along the
rotation, thumb along $\bm{T}$); with this orientation, the ring drifts along $+\bm{B}$, as in all
our simulations ($\beta > 0$). (b)--(f) Simulation (Barkley model, $a = 1.1$, $b = 0.17$,
$\epsilon = 0.02$; initial ellipse with semi-axes 8 and 5). (b) Excitation wave (surface
$u = 1/2$) and filament (tube) in the saddle phase at $t = 12$; the quadrant of the wave
facing the viewer is cut away to reveal the filament. (c)--(f) Filament over one axis-switching
cycle, seen from the same direction: planar ellipse with its long axis along $x$ ($t = 4$),
saddle ($t = 12$), planar ellipse with its long axis along $y$ ($t = 20$), and saddle with
reversed orientation ($t = 28$). The filament is colored by its height relative to the ring
center, in space units (red: above, blue: below); the gray curve is its projection onto the
horizontal plane. The bar in (f) is 5 space units long and applies to (c)--(f). Panels (b)--(f) are shown in a frame that moves with the ring center. The movie
shows panel (b) from $t = 3$ to 40, with the filament projected onto a horizontal plane below the
wave. (Multimedia available online.)}
\end{figure*}

For filaments whose radius of curvature is large compared with the size of the spiral
core---more precisely, with the width of the response functions of the spiral---this motion follows
a simple kinematic law.\cite{Keener1988,Biktashev1994} Each point of the filament moves with
a velocity that is proportional to the local curvature $K$ and that has two components,
\begin{equation}
  \frac{\partial \bm{X}}{\partial t} = \alpha K \bm{N} + \beta K \bm{B} ,
  \label{eq:kinematic}
\end{equation}
where $\bm{X}$ is the position on the filament, and $\bm{N}$ and $\bm{B}$ are the unit
normal and binormal vectors [Fig.~\ref{fig:cycle}(a)]. The normal vector points toward the
local center of curvature, and the binormal vector is perpendicular to the plane in which the
filament curves. The coefficient $\alpha$ is the filament tension. For $\alpha > 0$, a circular
filament of radius $R$ contracts as $\mathrm{d}R/\mathrm{d}t = -\alpha/R$, while for $\alpha < 0$, rings expand and the medium can develop scroll wave
turbulence.\cite{Biktashev1998,Alonso2003,Alonso2004,Bansagi2007} The coefficient $\beta$ moves a
circular ring along its symmetry axis with the velocity $\beta/R$; however, it vanishes if all reacting species
diffuse equally fast.\cite{Biktashev1994,BiktashevReview2014}

Equation~(\ref{eq:kinematic}) is closely related to a classical result of fluid dynamics. A
thin vortex filament in an ideal fluid---for example, the core of a smoke ring---moves in the
local induction approximation (LIA) with the velocity $\beta K \bm{B}$, that is,
Eq.~(\ref{eq:kinematic}) with $\alpha = 0$.\cite{Hasimoto1972,Saffman1992} J.~J. Thomson calculated
the vibrations of a slightly deformed vortex ring in 1883 and found that an elliptic deformation
oscillates, so that the long and short axes of the ring repeatedly exchange
places.\cite{Thomson1883} Arms and Hama followed this oscillation for initially planar, elliptic
rings in the LIA through the saddle shape and back to a planar ellipse with interchanged
axes.\cite{ArmsHama1965} Dhanak and De Bernardinis computed it
with the cut-off approximation of the Biot--Savart law\cite{Crow1970} and observed it in experiments
with air puffed through elliptic orifices; the motion was periodic for slightly elliptic
rings, and strongly elongated rings split into two.\cite{Dhanak1981} Viscous elliptic vortex rings and elliptic
jets show a similar exchange of axes, known as axis switching.\cite{Oshima1988,Cheng2016,LiHong2024,HussainHusain1989}

For scroll waves, the tension term adds dissipation to this picture. A similar combination of
binormal and dissipative normal motion describes vortices in superfluid
helium.\cite{Schwarz1985,VanGorder2015} In the complex Ginzburg--Landau equation for oscillatory media,
however, untwisted filaments have no binormal motion at lowest order and simply
contract.\cite{Gabbay1997,Aranson2002}

In excitable media, Keener and Tyson described noncircular filaments in thin BZ layers by purely
normal, curvature-driven motion,\cite{KeenerTyson1988} and Dutta and Steinbock showed that
planar, hairpin-shaped filaments in the BZ reaction and in the Barkley model with equal diffusion
coefficients translate with constant shape, as predicted by the curve-shortening
flow.\cite{Dutta2010} In these cases, the binormal motion is negligible
and the filament stays in its plane. In a BZ gel, Winfree and Jahnke photographed an initially elliptic scroll ring
that became round and then broke at two opposite sides, which they attributed to the filament reaching the
gel boundary where its curvature had been largest.\cite{WinfreeJahnke1989}

Knotted and linked filaments, anticipated by Winfree and
Strogatz,\cite{WinfreeStrogatz1984} preserve their topology and can persist for thousands of
rotation periods,\cite{Sutcliffe2003,Binysh2019} whereas tangled unknots relax toward
circles.\cite{Maucher2016} Such structures recall Kelvin's 1867 theory of vortex atoms, in which atoms
are vortex rings in a frictionless cosmic fluid, possibly linked or knotted, with modes of vibration
that give rise to spectral lines.\cite{Kelvin1867}

B\'ans\'agi and Steinbock analyzed the growth of the shape modes of
expanding BZ scroll rings under negative tension,\cite{Bansagi2007} but the oscillation that the
binormal term drives in such modes has not been studied. Moreover, Eq.~(\ref{eq:kinematic}) is only
the lowest order of a more general theory that includes curvature-dependent coefficients, a
fourth-order term known as filament rigidity, and couplings to the twist of the
filament.\cite{Dierckx2012,Dierckx2013,Nakouzi2014}

Here, we use three-dimensional simulations of the Barkley model\cite{Barkley1991} to study
scroll rings whose filaments start out as planar ellipses. We show that these rings perform the
axis-switching oscillation of elliptic vortex rings for vanishing, positive, and negative filament
tension, with a period that scales approximately as predicted by Eq.~(\ref{eq:kinematic}). At our
main parameters, however, the oscillation is about 1.45 times slower than predicted from the drift
of circular rings of the same size, which we trace to the curvature dependence of the drift
coefficient and to a weaker coupling of out-of-plane to in-plane
deformations. Two further models of excitable media show the same oscillation
with frequencies that differ from the prediction by about 6\% or less.

\section{Elliptic rings in the kinematic picture}
\label{sec:theory}

The binormal term moves each part of the filament perpendicular to its local plane of curvature,
with a speed proportional to the curvature. For an initially planar ellipse, this motion is along
the ring axis and is fastest at the two ends of the long axis. These ends therefore advance ahead
of the flatter sides, deforming the ring into a saddle [Fig.~\ref{fig:cycle}(b,d)]. The leading
ends now form crests, while the lagging sides form troughs. This out-of-plane bending tilts the
binormal velocity inward at the leading ends and outward at the lagging sides. Accordingly, the
original long axis shortens while the short axis lengthens, until the ring becomes planar again
with its axes interchanged [Fig.~\ref{fig:cycle}(e)]. The cycle then repeats. The tension term
damps the oscillation for $\alpha > 0$ and amplifies it for $\alpha < 0$.

To make this picture quantitative, we consider a ring that deviates slightly from a circle of
radius $R$. At the polar angle $\theta$, its radial deformation is $r(\theta)$ and its vertical
deformation is $\zeta(\theta)$. We expand both in Fourier modes $\cos(m\theta)$, with amplitudes
$r_m$ and $\zeta_m$. The mode $m = 1$ shifts or tilts the ring, and $m = 2$ makes it elliptic.
Linearizing Eq.~(\ref{eq:kinematic}) yields (see the supplementary material for this and the
following derivations)
\begin{align}
  R^2\,\dot r_m &= -\alpha\,(m^2-1)\,r_m - \beta\, m^2\, \zeta_m , \nonumber\\
  R^2\,\dot \zeta_m &= \beta\,(m^2-1)\,r_m - \alpha\, m^2\, \zeta_m .
  \label{eq:modes}
\end{align}
The binormal terms couple the in-plane and out-of-plane deformations; the resulting exchange is the
axis-switching oscillation. To linear order, the deformations average to zero around the ring and hence do
not affect the rates of change of the mean radius and the mean height. Accordingly, the ring drifts like a circle of the same mean
radius, with the velocity $\beta/R$, while its shape oscillates. For $\alpha = 0$, the mode
oscillates with the angular frequency
\begin{equation}
    \omega_m = \frac{|\beta|\, m\sqrt{m^2-1}}{R^2}.
  \label{eq:omega}
\end{equation}
Successive axis switches of an ellipse ($m = 2$) are separated by the half-period
$T_{1/2} = \pi R^2/(2\sqrt{3}\,|\beta|)$. An initially planar ellipse reaches its first switch after
$T_{1/2}/2$, when its projection onto the original plane is circular, and becomes planar with interchanged axes after $T_{1/2}$. The full period equals $1/(2\sqrt3) = 0.289$ times the time
$2\pi R^2/|\beta|$ that the ring needs to travel its own circumference, in agreement with Thomson's
result.\cite{Thomson1883} Between successive switches, the ring hence drifts along its axis by
$\pi R/(2\sqrt3) = 0.91\,R$, independent of $\beta$.

The eigenvalues of the system of Eqs.~(\ref{eq:modes}) are
$\lambda = [-\alpha(2m^2-1) \pm \sqrt{\alpha^2 - 4\beta^2 m^2(m^2-1)}\,]/(2R^2)$; hence an
elliptic ring ($m = 2$) oscillates only if $|\beta|/|\alpha| > 1/(4\sqrt{3}) \approx 0.14$. For
$\alpha > 0$, it then shrinks while it oscillates with decaying amplitude; for $\alpha < 0$, it
expands while the deformation grows.

Because both $\mathrm{d}\ln R/\mathrm{d}t$ and the oscillation frequency scale with $1/R^2$, a ring
with constant coefficients oscillates at a constant rate in the logarithm of its radius, which means
faster and faster in time if it shrinks. For the elliptic mode, the phase advances by $\Omega/|\alpha|$,
with $\Omega = \sqrt{48\beta^2-\alpha^2}/2$, for every factor $e$ by which the radius changes, and the
envelope of the relative ellipticity $r_2/R$ varies as $(R/R_0)^{5/2}$. The number of axis switches per
factor $e$ is hence $\Omega/(\pi|\alpha|)$, which depends only on $\beta/\alpha$. A shrinking ring with
$|\beta| \gg \alpha$ switches its axes many times before collapse, whereas a ring with
$|\beta| \approx \alpha$ switches about once per factor $e$.

Equations~(\ref{eq:modes}) and (\ref{eq:omega}) assume constant $\alpha$ and $\beta$. To
discuss deviations from this lowest-order picture, we write the linearized mode equations more
generally as
\begin{align}
  R^2\,\dot r_m &= -\alpha_r\,(m^2-1)\,r_m - \beta_\zeta\, m^2\, \zeta_m , \nonumber\\
  R^2\,\dot \zeta_m &= \beta_r\,(m^2-1)\,r_m - \alpha_\zeta\, m^2\, \zeta_m ,
  \label{eq:modes2}
\end{align}
where $\beta_\zeta$ converts vertical into radial deformations and $\beta_r$ converts radial into
vertical ones, and $\alpha_r$ and $\alpha_\zeta$ damp or amplify the two deformations. If the coefficients
depend on the local curvature alone, the two couplings are affected differently: a radial deformation changes the
local curvature, so that
\begin{equation}
  \beta_r = \frac{\mathrm{d}(\beta K)}{\mathrm{d}K} = \beta + K\,\frac{\mathrm{d}\beta}{\mathrm{d}K} ,
  \label{eq:betaeff}
\end{equation}
whereas a vertical deformation only tilts $\bm{N}$ and $\bm{B}$ without changing the curvature
to first order, so that $\beta_\zeta$ equals the drift coefficient $\beta_\mathrm{drift}$ of the
ring for every $m$. Analogously,
$\alpha_r = \mathrm{d}(\alpha K)/\mathrm{d}K$ and $\alpha_\zeta = \alpha$ in such a curvature-only
law. Higher-order local laws change these relations: the rigidity terms
$-(e_1 + e_2\,\bm{T}\times)(\partial_s^4\bm{X})_\perp$ of Dierckx \textit{et al.}\cite{Dierckx2012}
add $(m^2-1)\,e_2K^2$ to $\beta_\zeta$ and $m^2 e_2K^2$ to $\beta_r$, and analogous terms with $e_1$
to $\alpha_\zeta$ and $\alpha_r$ (supplementary material).

An oscillating solution of Eqs.~(\ref{eq:modes2}) decays at the rate
$\gamma = [\alpha_\zeta m^2 + \alpha_r(m^2-1)]/(2R^2)$ and has the frequency
$\omega_m^2 = m^2(m^2-1)\beta_\zeta\beta_r/R^4 - \Delta^2/4$, where
$\Delta = [\alpha_\zeta m^2 - \alpha_r(m^2-1)]/R^2$ is the difference of the two diagonal rates. The
amplitudes of the two deformations are related by
\begin{equation}
  \frac{|\zeta_m|}{|r_m|} = \sqrt{\frac{m^2-1}{m^2}}\,\sqrt{\frac{\beta_r}{\beta_\zeta}} ,
  \label{eq:ampratio}
\end{equation}
where $|\zeta_m|$ and $|r_m|$ are the amplitudes of the two damped sinusoids.
Equation~(\ref{eq:ampratio}) holds for any diagonal rates; $\Delta$ only shifts the radial and
vertical oscillations away from quadrature, by an angle $\phi$ with $\tan\phi = \Delta/(2\omega_m)$.
Frequency, amplitude ratio, decay rate, and phase therefore yield all four coefficients; numerical solutions
of Eq.~(\ref{eq:kinematic}) with a curvature-dependent $\beta$ agree with the predicted frequencies and
amplitude ratios within 0.6\% and 1.5\% (supplementary material).

\section{Model and methods}
\label{sec:model}

For most of our numerical experiments, we use the Barkley model\cite{Barkley1991}
\begin{align}
  \frac{\partial u}{\partial t} &= \nabla^2 u + \frac{1}{\epsilon}\, u(1-u)\left(u - \frac{v+b}{a}\right), \nonumber\\
  \frac{\partial v}{\partial t} &= u - v ,
  \label{eq:barkley}
\end{align}
in which $u$ is a fast activator and $v$ a slow inhibitor that does not diffuse. We fix
$\epsilon = 0.02$. Our main parameter set is $a = 1.1$, $b = 0.17$, for which two-dimensional
spirals rotate rigidly with a period of 4.44 around a circular core of radius 0.49 and emit
waves of wavelength 12.0. Additional simulations use $a = 1.1$ with $b = 0.15$, 0.16, and 0.18
(also rigid, with periods of 4.10 to 4.69, core radii of 0.37 to 0.58, and wavelengths of 11.2 to
12.7; supplementary material), $a = 0.8$ with $b = 0.05$ (rigid), and $a = 0.8$ with $b = 0.09$, for
which spirals meander.

Equations~(\ref{eq:barkley}) are solved with an explicit Euler scheme on a cubic grid of
spacing $\Delta x = 0.2$, a seven-point Laplacian, and the time step
$\Delta t = 0.8\,\Delta x^2/6$, in periodic domains in which the ring stays at least one wavelength
from its periodic images and is kept at the center by vertical shifts of the fields. Larger domains,
a halved time step, and a seed turned by $45^\circ$ relative to the grid change the mean frequency
ratio of Sec.~\ref{sec:period} by less than 1\%; grid spacings of 0.15 and 0.1 change it by 2 to
3\% (supplementary material).

Scroll rings are initiated by a broken wave confined to the interior of an ellipse with
semi-axes $A_x$ and $A_y$ (written $A_x \times A_y$ below): inside the ellipse, a thin excited
layer ($u = 1$) is placed below a refractory layer ($v = 1$). The free edge of this wave curls up
and forms a single scroll ring within about three time units. Rings with an $m$-fold deformation are
prepared with the boundary $r = A(1 + \delta\cos m\theta)$, where $\delta = 0.08$ to 0.15, and tilted
rings by rotating the plane of a circular seed by $12^\circ$. We also simulate straight scroll waves
whose filament is initially displaced by $0.5\cos(kz)$ in a domain that is periodic in $z$ with
height $2\pi/k$ (supplementary material).

The filament is computed as the intersection of the isosurfaces $u = 1/2$ and
$v = a/2 - b$.\cite{Barkley1991} This curve follows the spiral tip, which circles the center of
rotation once per rotation period; we remove this wobble by averaging all ring measures over one
period. We analyze only frames in which the extracted filament is a single closed curve (more
than 99\% of all frames). From the filament we compute its length $L$ and $\bar R = L/2\pi$; the
arc-length-weighted mean horizontal distance $R_m$ of the filament from its center; the
half-widths $h_x$ and $h_y$ of the ring along $x$ and $y$; the axial position of its center; and
the Fourier amplitudes $r_m$ and $\zeta_m$ of its horizontal distance from the center and of its
height, as functions of the polar angle $\theta$ and weighted by angle. An axis switch occurs whenever $r_2$
changes sign, and $\zeta_2$ is the saddle amplitude; a half-cycle is the interval between two
switches. We denote by $\beta_\mathrm{c}(R)$ and $\alpha_\mathrm{c}(R)$ the
coefficients of circular rings of radius $R$ (Sec.~\ref{sec:radius}) and by
$\beta_\mathrm{own} = \bar R v$ the drift coefficient that a deformed ring has according to its own
drift speed $v$; for the weakly deformed rings of Sec.~\ref{sec:modes}, we use
$\beta_\mathrm{drift} = R_m v$, which differs from $\beta_\mathrm{own}$ by at most 1.4\%.
Alternative filament definitions and the numerical solution of Eq.~(\ref{eq:kinematic}) are described
in the supplementary material.

\textit{Use of generative AI.} This study made extensive use of the large language model
Claude Opus 5.5 (Anthropic), accessed in October and November 2026 through the Claude Code desktop
application. Under the author's direction, the model wrote and debugged the simulation and analysis
code, executed the simulations, analyzed the results and contributed to their interpretation, derived
the linear mode analysis of Sec.~\ref{sec:theory}, produced
the figures with plotting scripts, searched the literature, and drafted the manuscript text. To
identify errors, Claude agents, ChatGPT 6.1-Sol, and GPT-6 Astra Pro (both OpenAI) reviewed the deposited
code, the manuscript, and the supplementary material. All data were produced by the deposited, deterministic
simulation code, and all figures are plots generated from these data by deposited scripts. The author
critically reviewed all AI-generated content.

\section{Numerical results}
\label{sec:results}

\subsection{Axis switching}
\label{sec:switching}

Figure~\ref{fig:cycle} (Multimedia available online) shows a scroll ring that starts from an
$8 \times 5$ ellipse at $a = 1.1$, $b = 0.17$. After its formation, the filament is a nearly planar
ellipse with half-widths of 6.8 and 4.2 at $t = 4$ [Fig.~\ref{fig:cycle}(c)]. The ends of the long
axis, where the filament is most strongly curved, immediately move ahead of the rest of the ring,
and the ring buckles into a saddle [$t = 12$, Fig.~\ref{fig:cycle}(b,d)]. At $t \approx 20$, the ring is
planar again, but now its long axis points in the $y$-direction; at $t \approx 37$, it has returned to its
original shape. Meanwhile, the ring travels steadily along its axis by about 0.33 space units
per time unit.

\begin{figure}
\includegraphics[width=\columnwidth]{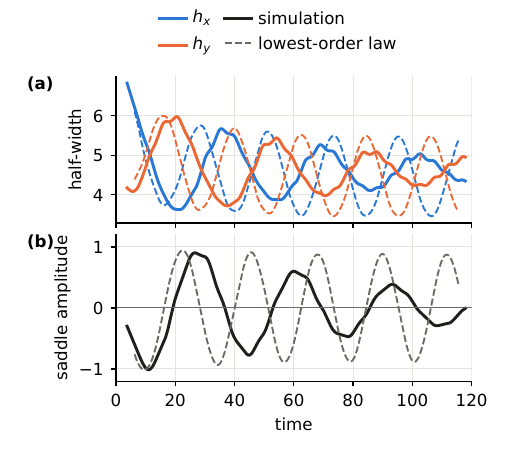}
\caption{\label{fig:smallring} Axis switching of the elliptic ring of Fig.~\ref{fig:cycle}.
(a) Half-widths $h_x$ (blue) and $h_y$ (orange) of the ring. (b) Saddle amplitude $\zeta_2$. Curves:
reaction--diffusion simulation (solid) and the lowest-order law (dashed; see text), both averaged over one
rotation period.}
\end{figure}

As shown in Fig.~\ref{fig:smallring}, the oscillation is long-lived. Over 120 time units, the axes
switch seven times while the amplitude decreases by a factor of 3 to 4. We compare the simulated ring with what we refer to as the lowest-order law, namely
Eq.~(\ref{eq:kinematic}) with the coefficients of circular rings of the same size. For this ring,
we set $\alpha = 0$, start from the measured shape at $t = 4$, and let the model ring follow the measured
filament length of the simulated ring, so that both rings always have the same size (dashed). This law
captures the shape of the cycle, but its half-cycles are 1.37 to 1.46 times shorter than those of the
simulated ring, and its oscillation does not decay. Following the measured length matches the size of the
two rings, not their shape, so that the law keeps its relative deformation whereas the simulated ring
becomes rounder. We quantify the difference in period half-cycle by half-cycle in Sec.~\ref{sec:period}.

\begin{figure}
\includegraphics[width=\columnwidth]{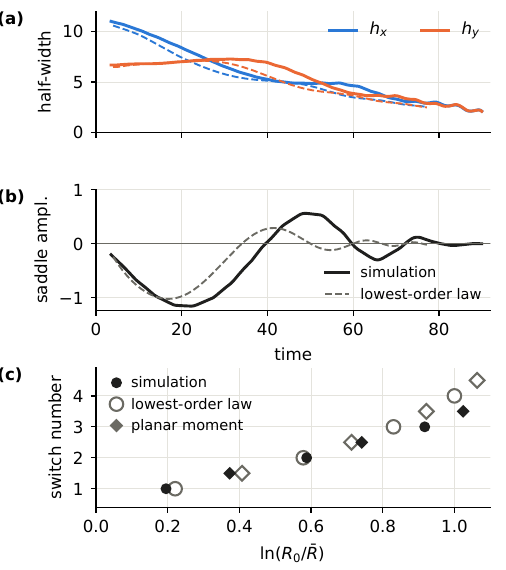}
\caption{\label{fig:shrinking} Axis switching of a shrinking ring under positive tension ($a = 1.1$,
$b = 0.15$, initial $12 \times 7.5$ ellipse). (a) Half-widths and (b) saddle amplitude; solid:
simulation; dashed: lowest-order law, Eq.~(\ref{eq:kinematic}) with the tension and drift
coefficient of circular rings of the same instantaneous size (Fig.~S1 of the supplementary
material), started from the measured shape at $t = 4$ and shown until the model ring radius drops
below 2.5. (c) Number of axis switches (circles) and planar moments midway between them (diamonds), from the crossings in (a) and (b), versus the logarithmic decrease of the ring radius, $\bar R
= L/2\pi$, for the simulation (filled) and the lowest-order law (open), shown while $\bar R > 3$.}
\end{figure}

Axis switching is not restricted to rings with vanishing tension. At $b = 0.15$, the tension is
clearly positive, and an initially $12 \times 7.5$ ellipse shrinks from a mean radius of 9 to
collapse within 92 time units while switching its axes four times (Fig.~\ref{fig:shrinking}).
The intervals between switches shorten from 26 to about 9 time units, but they stay nearly
constant on a logarithmic radius scale [0.39 and 0.33 in $\ln\bar R$, Fig.~\ref{fig:shrinking}(c)],
as expected from Sec.~\ref{sec:theory}. The lowest-order law with the coefficients of circular
rings of the same size places the first two switches at similar values of $\ln\bar R$ but reaches
them earlier, because it shrinks the ring faster; the positions in $\ln\bar R$ test only $\beta/\alpha$. For $a = 0.8$,
$b = 0.05$, where the shrinkage and drift of the elliptic ring give $\alpha = 1.35$ and
$|\beta| = 1.02$ ($|\beta|/\alpha \approx 0.75$), the ring completes only a single, strongly
damped switch before it collapses, again in line with the linear theory. At $b = 0.18$, the tension
is negative; the ring grows from a mean radius of 8.5 to 12 and switches its axes twice within 90
time units while its relative ellipticity grows to about 0.4. The switching also occurs for the meandering
spiral at $a = 0.8$, $b = 0.09$.

\subsection{Tension and drift of circular rings}
\label{sec:radius}

\begin{figure*}
\includegraphics[width=\textwidth]{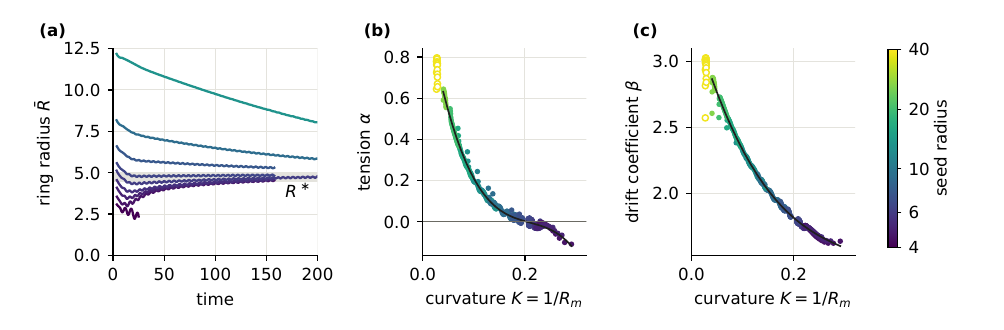}
\caption{\label{fig:circles} Circular scroll rings at $a = 1.1$, $b = 0.17$. (a) Ring radius
versus time for seed radii between 4 and 13; the gray band marks the final radii of the rings that
approach $R^\ast \approx 4.5$--4.9 from below and above. (b) Tension $\alpha$ and (c) drift coefficient $\beta$ versus
curvature $K = 1/R_m$, measured in 10-unit time windows after $t = 20$ on 14 rings with seed radii of 4.5 to 26;
colors encode the seed radius, and the lines are the polynomial fits used in the analysis. Open
circles: early windows of a ring with seed radius 38, which develops undulations and is not used.}
\end{figure*}

To predict the oscillation period independently of the elliptic rings, we measure the tension and
drift coefficient of circular rings from their contraction, $\alpha = -\tfrac{1}{2}\,\mathrm{d}R_m^2/\mathrm{d}t$, and their drift,
$\beta = \mathrm{d}z/\mathrm{d}s$ with $\mathrm{d}s = \mathrm{d}t/R_m$ (Fig.~\ref{fig:circles}). Both
coefficients have to be measured, because large-excitability asymptotics predicts vanishing $\alpha$
and $\beta$ at leading order for the Barkley model [Ref.~\onlinecite{Margerit2002a}, Secs.~III~A and
IV~D; Ref.~\onlinecite{Margerit2002b}, Sec.~VI]. For $a = 1.1$, $b = 0.17$, rings of very different
sizes collapse onto single curves [Fig.~\ref{fig:circles}(b,c)], and both coefficients depend
strongly on curvature. As $K$ increases from 0.04 to 0.29, the tension falls from 0.63 through zero
at $K = 0.20$ to $-0.12$, while the drift coefficient falls from 2.9 to 1.6. At small $K$, $\beta$
falls steeply, more like a logarithm than like the polynomial in $K^2$ expected from a local gradient
expansion\cite{Dierckx2013} (supplementary material). Polynomial fits to these curves provide $\alpha_\mathrm{c}(R)$
and $\beta_\mathrm{c}(R)$ for the analysis below, and we calibrate $b = 0.15$, 0.16, and 0.18 in the
same way (supplementary material).

Because the tension changes sign [Fig.~\ref{fig:circles}(b)], large rings shrink and small rings grow: rings seeded with radii
of 4.5 to 6.5 approach a nearly stationary radius from both sides and end between $\bar R = 4.53$
and 4.82 while still converging slowly [gray band in Fig.~\ref{fig:circles}(a)], whereas a ring
seeded with radius 4 collapses. We regard this range, $R^\ast \approx 4.5$--4.9, as a long-lived,
nearly stationary radius rather than as an established attractor (our runs cover at most 300 time
units, and $R^\ast$ moves to about 4.5 at $\Delta x = 0.15$); here, it provides a background state of
nearly constant size for the mode analysis of Sec.~\ref{sec:modes}. At
$b = 0.16$, rings settle at $R^\ast \approx 3.1$, whereas at $b = 0.15$ all rings collapse and at
$b = 0.18$ all rings expand. Scroll rings whose contraction stops at a finite radius were reported
earlier in simulations;\cite{Brazhnik1987,Skaggs1988,Courtemanche1990} whether this reflects
higher-order terms of the filament law or an interaction between opposite parts of the ring is an
open question.\cite{BiktashevReview2014}

\subsection{Oscillation period}
\label{sec:period}

Next, we test whether the lowest-order law predicts the axis-switching period quantitatively. The
coefficients of circular rings (Sec.~\ref{sec:radius}) leave this comparison without any adjustable
parameter. For every half-cycle between two axis switches, we integrate the angular frequency that
Eqs.~(\ref{eq:modes}) predict with the coefficients $\alpha_\mathrm{c}$ and $\beta_\mathrm{c}$ of
circular rings of the measured radius $\bar R(t)$ over the observed duration $T_{1/2}$ and divide by
$\pi$. As rings shrink by up to 32\% within one half-cycle, we integrate over $\bar R(t)$ rather than
use a single radius. The result is the ratio of the predicted to the observed mean frequency. This
frequency ratio states how many times slower the ring oscillates than predicted. The prediction
includes two small corrections: the nonlinear Eq.~(\ref{eq:kinematic}) changes the half-period by at
most 2\%, and the tension changes the frequency by less than 0.2\%. Figure~\ref{fig:period}(a) shows the result for 22
half-cycles from five simulations (all values in the supplementary material).

At $b = 0.17$, 13 half-cycles of two strongly elliptic rings with $\bar R = 4.6$ to 8.0 give
frequency ratios of 1.41 to 1.49 (mean 1.450), without a trend within either run, and weakly deformed
rings at $R^\ast$ (Sec.~\ref{sec:modes}) give $1.48 \pm 0.03$. The ratio hence depends on the
amplitude of the deformation by only a few percent. At $b = 0.15$ and 0.16, the ratio is 1.20 to
1.43, and at $b = 0.18$, the single analyzed half-cycle gives 1.47.
We use $\beta_\mathrm{c}$ rather than the ring's own, smaller drift coefficient $\beta_\mathrm{own}$
(Sec.~\ref{sec:model} and supplementary material), because $\beta_\mathrm{c}$ is measured independently
of the elliptic rings. With $\beta_\mathrm{own}$, the ratios would be smaller ($1.33 \pm 0.07$ at
$b = 0.17$) and would vary within each run. For the meandering spiral,\cite{Dierckx2017} averaging over one rotation period does
not remove the meander, and we therefore do not analyze its period.

\begin{figure*}
\includegraphics[width=\textwidth]{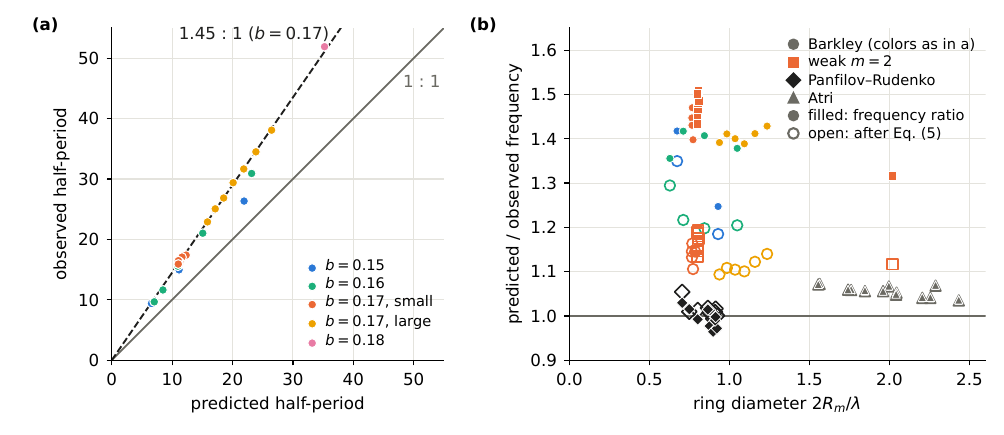}
\caption{\label{fig:period} (a) Observed versus predicted half-period of the axis-switching
oscillation for 22 half-cycles of five simulations at $a = 1.1$ ($b = 0.17$ small: seed $8\times 5$,
Fig.~\ref{fig:smallring}; $b = 0.17$ large and $b = 0.15$, 0.16, 0.18: seeds $12\times 7.5$), predicted
with the coefficients of circular rings of the measured radius $\bar R(t) = L/2\pi$
(Sec.~\ref{sec:period}). Solid line: equality; dashed line: mean ratio at $b = 0.17$.
(b) Frequency ratio (filled) and the same ratio divided by the slowdown that the curvature-only law,
Eq.~(\ref{eq:betaeff}), predicts (open; the two coincide for the Atri model) for half-cycles between planar moments, versus the ring
diameter $2R_m$ in spiral wavelengths: Barkley rings of (a) except $b = 0.18$ (colors as in a), weakly deformed Barkley
rings at $R^\ast$ and at a diameter of 2.0 wavelengths (squares; the latter from the mode fit of Sec.~\ref{sec:modes}), the
Panfilov--Rudenko model (diamonds), and the Atri model (triangles). The two Barkley
half-cycles with $\bar R < 3.5$, where $\beta(K)$ is poorly constrained, are omitted.}
\end{figure*}

The curvature dependence of the drift coefficient accounts for much, but not all, of the observed slowdown. In
Eqs.~(\ref{eq:modes2}), the oscillation frequency depends mainly on $\sqrt{\beta_r\beta_\zeta}$. A
curvature-only law keeps $\beta_\zeta = \beta_\mathrm{c}$, whereas Eq.~(\ref{eq:betaeff}) with the
measured $\beta(K)$ gives $\beta_r \approx 0.63\,\beta_\mathrm{c}$ at $b = 0.17$, which lengthens the
predicted period by a factor of $\sqrt{\beta_\mathrm{c}/\beta_r} \approx 1.26$. For the weakly
deformed rings at the nearly stationary radius $R^\ast$, where linear theory applies, this factor
accounts for most of the observed slowdown of 1.48 and leaves a residual of about 1.17 that no
curvature-only law explains. For the strongly elliptic rings, the factor is 1.19 to 1.27 and the
residual 1.09 to 1.23, depending on how the factor is evaluated (supplementary material). At
$b = 0.15$ and 0.16, where $\beta$ depends more weakly on curvature, the factor is 1.0 to 1.25 and
the residual 1.1 to 1.4. The observed half-periods range from 9.4 to 52 and approximately follow the
$\bar R^2/|\beta|$ scaling, with a prefactor that depends on $b$.

\subsection{Shape modes}
\label{sec:modes}

\begin{figure}
\includegraphics[width=\columnwidth]{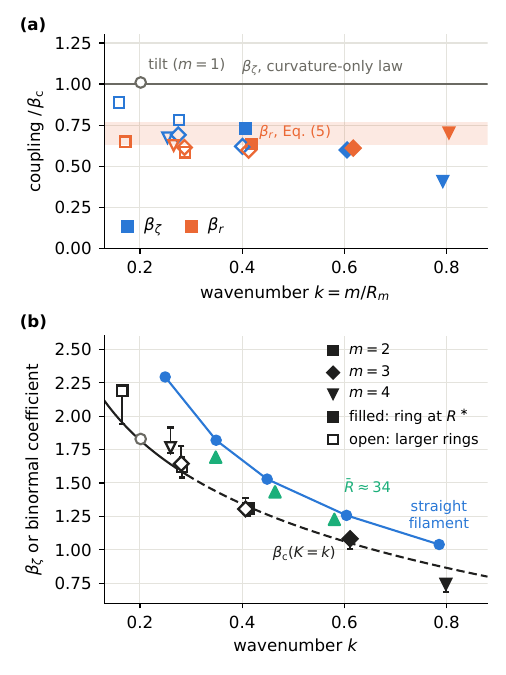}
\caption{\label{fig:betaratio} Response of scroll wave filaments to deformations of wavenumber
$k$ ($a = 1.1$, $b = 0.17$). (a) Couplings $\beta_\zeta$ (vertical to radial) and $\beta_r$
(radial to vertical) of ring modes relative to $\beta_\mathrm{c}$ of circular rings of the same radius,
versus $k = m/R_m$: twofold (squares), threefold (diamonds), and fourfold (triangles) modes at
$R^\ast$ (filled) and on larger rings (open), with the two couplings offset by $\mp 0.006$ in $k$ for
visibility. The gray line marks $\beta_\zeta$ of the curvature-only law, the shaded band the range of
$\beta_r$ that this law predicts through Eq.~(\ref{eq:betaeff}), and the gray circle
$\beta_\zeta/\beta_\mathrm{drift}$ of the tilted ring ($m = 1$), which equals 1 by symmetry. (b) Absolute
$\beta_\zeta$ of the ring modes (bars: ranges of Table~S5; gray circle: tilted ring); $\beta_\mathrm{c}$
of circular rings with $K = k$ (black line; dashed: extrapolated); the binormal coefficient of straight
filaments with a small sinusoidal deformation (blue; bars: uncertainty of the extrapolation to zero
amplitude); and the coefficient obtained from the frequencies of the growing undulations of a ring with
initial radius 38 ($\bar R \approx 34$; green). Fig.~S3 also shows the data at $b = 0.15$ and 0.16.}
\end{figure}

The remaining slowdown comes mainly from the out-of-plane coupling $\beta_\zeta$, which depends on the
wavenumber $k = m/R$ of the deformation rather than on the curvature of the ring. To show this, we
prepared weakly deformed rings with twofold, threefold, and fourfold modes at $R^\ast$ and on larger
rings with radii of up to about 15. Damped sinusoids fitted to the radial and vertical mode amplitudes
give the four coefficients of Eqs.~(\ref{eq:modes2}) (supplementary material, Tables~S5, S6, and S11).
A curvature-only law requires $\beta_\zeta = \beta_\mathrm{c}$, but $\beta_\zeta$ falls with increasing
$k$, to $0.41\,\beta_\mathrm{c}$ for $m = 4$ at $R^\ast$ [Fig.~\ref{fig:betaratio}(a)]. Rings with
different $m$ and $R$ but equal $k$ nevertheless have nearly the same $\beta_\zeta$: 1.31 for both
rings at $k \approx 0.41$, and 1.62 and 1.64 at $k \approx 0.28$ [Fig.~\ref{fig:betaratio}(b)]. In
contrast, $\beta_r$ lies within 19\% of Eq.~(\ref{eq:betaeff}), and within 4\% for $m = 2$ and 3 at
$R^\ast$. The amplitude ratio of Eq.~(\ref{eq:ampratio}) separates the two couplings directly: for
$m = 2$ at $R^\ast$, it is 0.81, against 0.69 for the curvature-only law.

At $b = 0.17$, $\beta_\zeta$ falls close to the drift coefficient of a circular ring whose curvature
equals $k$: all values lie within 14\% of $\beta_\mathrm{c}(K = m/R)$, including three rings that we
prepared to test this relation. Note that for $k > 0.29$, this comparison extrapolates
$\beta_\mathrm{c}$ beyond its calibrated range. For $m = 2$ at $R^\ast$, the reduction slows the
oscillation by $\sqrt{\beta_\mathrm{c}/\beta_\zeta} = 1.17$, matching the residual of 1.17
(Sec.~\ref{sec:period}). However, this relation does not carry over to other parameters. At
$b = 0.15$, $\beta_\mathrm{c}$ varies little with curvature, yet $\beta_\zeta$ still falls with $k$ and
stays within about 16\% of its values at $b = 0.17$ (Fig.~S3); the same holds at $b = 0.16$. It is hence likely
that the agreement with $\beta_\mathrm{c}$ at $b = 0.17$ is partly coincidental. At $b = 0.15$ and
0.16, $\beta_\zeta$ is about 0.7 to 0.85 of $\beta_\mathrm{c}$, which accounts for a residual of about
1.1 to 1.2, the lower part of the observed range.

Straight filaments with a small sinusoidal deformation show the same dependence. Their binormal
coefficient $\omega/k^2$ falls from 2.29 at $k = 0.25$ to 1.04 at $k = 0.79$, far from linear in $k^2$
as a fourth-order law would require [Fig.~\ref{fig:betaratio}(b)], and it barely changes between
$b = 0.15$ and 0.17 (Fig.~S3). At equal $k$, $\beta_\zeta$ of the ring modes lies about 20\% below
these values, whereas the growing undulations of a large ring ($\bar R \approx 34$) agree with them,
so that the ring curvature matters as well. The effect is not tied to the instability of straight
filaments at the main parameters,\cite{Dierckx2012} because straight filaments are stable at
$b = 0.15$.

The results at $R^\ast$ hold for all five filament definitions and at $\Delta x = 0.15$, and a fourfold
mode rotated by $22.5^\circ$ relative to the grid agrees within about 1\%. The tilted ring, for which
$\beta_\zeta = \beta_\mathrm{drift}$ under any dynamics, gives 1.01. Neither the twist that the
deformation induces nor a single rigidity coefficient $e_2$ accounts for the reduction, although the
rigidity term also depends on $k$. The $e_2$ fitted to $\beta_\zeta$ grows in magnitude from 1.8 at
$k = 0.80$ to 10--13 at $k = 0.17$--0.28, and the local slope of $\omega/k^2$ versus $k^2$ grows in the
same way, so that no single $e_2$ applies even at the smallest $k$ studied. Moreover, an $e_2$ matched to the frequency
moves the amplitude ratio of the $m = 2$ mode at $R^\ast$ away from the measured value (supplementary
material). On small rings, the modes are also damped more strongly than the curvature-only law allows:
at $R^\ast$, this law lets the $m = 2$ and 3 modes grow at rates of about 0.01 and 0.03, whereas they
decay at 0.011 and 0.058, and the $m = 4$ mode at 0.12. On our largest rings, this difference reverses:
the $m = 3$ and 4 modes grow slowly, at about 0.002 to 0.005, where this law predicts decay.

\section{Two further models}
\label{sec:second}

To test whether axis switching and its period depend on the specific kinetics, we use the
piecewise-linear model of Panfilov and Rudenko\cite{PanfilovRudenko1987} (supplementary material).
Note that, as in the Barkley model, its recovery variable $v$ does not diffuse, so that $\beta$ need
not vanish. The model suits our purpose
because its scroll rings change from contraction to expansion as the excitability parameter $g_f$
decreases, while they keep drifting along their axis.\cite{PanfilovRudenko1987} Near this
transition, the tension vanishes but the drift does not.

For $g_f = 0.775$ and 0.8, spirals rotate rigidly with wavelengths of 37 and 34, circular rings have
slightly negative and weakly positive tension ($\alpha = -0.032 \pm 0.009$ and $0.086 \pm 0.005$),
and their drift coefficient changes little with curvature (supplementary material). Elliptic rings
perform the same cycle as in the Barkley model (Fig.~S4). Over 1600 time units, a ring at $g_f = 0.775$ switches its axes
four times without changing its size much, and a ring at $g_f = 0.8$ switches six times while it
shrinks to half its radius. To compare the three models on equal terms, we time their half-cycles by
the planar moments ($\zeta_2 = 0$) and use $R_m$ as the radius (supplementary material). The
frequency ratio is then $0.971 \pm 0.006$ at $g_f = 0.775$ (three half-cycles) and
$1.007 \pm 0.014$ at $g_f = 0.8$ (seven half-cycles of four runs at grid spacings of 0.6 to 0.3),
compared with 1.41 to 1.48 for the Barkley rings at $b = 0.17$ analyzed in the same way. The small
shortfall at $g_f = 0.775$ is consistent with the curvature dependence of $\beta$, for which
Eq.~(\ref{eq:betaeff}) predicts 0.970, although this factor depends on the grid spacing by a few
percent (supplementary material). The damping again departs from the constant-coefficient law: at
$g_f = 0.775$, the relative ellipticity at successive planar moments decreases from 0.21 to 0.12
while the ring grows slightly, whereas the slightly negative tension of circular rings would let it
grow by about 20\%. Axis switching is hence not a peculiarity of one model, whereas the size of the
departure from the lowest-order law depends on the kinetics: at comparable ring sizes in units of
the spiral wavelength [Fig.~\ref{fig:period}(b)], the Panfilov--Rudenko rings are consistent with
the curvature-only law, and the Barkley rings are not. Note that the larger relative size of the Atri
rings does not explain their agreement with the prediction, because a weakly deformed Barkley ring of
similar size still oscillates about 1.3 times slower than predicted [Fig.~\ref{fig:period}(b)]. One might suspect an interaction of opposite
sides of the ring through the emitted waves. However, Panfilov--Rudenko rings of similar diameter show no
residual, and the Barkley residual does not change between diameters of 0.8 and 1.3 wavelengths.
Nandapurkar and Winfree simulated a scroll ring
in this model at $g_f = 0.78$ that became elliptic and periodically interchanged its axes (their
Fig.~3), an effect they attributed to the walls of their box;\cite{Nandapurkar1989} we suspect that
it was the binormal axis switching described here, which our rings show without walls.

We also tested a model with a direct physiological counterpart, the model of Atri
\textit{et al.} for Ca$^{2+}$ waves in \textit{Xenopus} oocytes,\cite{Atri1993} in which Ca$^{2+}$
diffuses with $D_c = 20~\mu$m$^2$\,s$^{-1}$ and the fraction of IP$_3$ receptors that are not
inactivated by Ca$^{2+}$ does not diffuse. At a constant IP$_3$ concentration of 0.1~$\mu$M, spirals
rotate rigidly with a wavelength of 120~$\mu$m, and circular rings expand slowly with a drift
coefficient that is nearly independent of the radius (supplementary material). Elliptic rings with
mean radii between 93 and 146~$\mu$m again switch their axes (Fig.~S5) and oscillate
$5.7 \pm 1.2\%$ more slowly than predicted (13 half-cycles of five rings), or $4.5 \pm 0.8\%$ for the
five half-cycles that start from nearly circular rings. In this respect, the Atri model is close to
the Panfilov--Rudenko kinetics rather than to the Barkley model. Rings of 190 to 290~$\mu$m diameter
would switch their axes every 5 to 12~min, although such rings may be hard to form and observe in
oocytes (supplementary material).

We also considered simulations with the two-variable Oregonator model of the BZ reaction, for
immobilized catalysts and for diffusing ones such as Ce(III/IV). However, the regimes we examined did
not allow a test of axis switching. With an immobilized catalyst, spirals mostly meandered. With a
diffusing catalyst, spirals rotated rigidly, but rings contracted too quickly to switch their axes. Near
the transition to negative
tension,\cite{Alonso2006} contraction should be slow, but the large spiral core would require very large
rings and long runs, which we did not attempt (supplementary material).

\section{Conclusions}
\label{sec:discussion}

One and a half centuries ago, Thomson showed that elliptic vortex rings in fluid dynamics oscillate as they travel,
with their long and short axes repeatedly exchanging places.\cite{Thomson1883} We have found that elliptic scroll rings
in excitable media perform the same oscillation, although their motion involves no macroscopic mass
transport and arises from reaction and diffusion instead of fluid inertia. The oscillation occurred for positive, negative, and vanishing filament tension, and between
two axis switches, the ring must travel about one radius.

The filament law that also describes vortex rings predicts the period of this oscillation without any
adjustable parameter, and two of our three models kept this time within about 6\%. The Barkley model,
however, ran about 1.45 times slower at our main parameters. Elliptic scroll rings hence act as
sensitive clocks whose rate reflects the kinetics of the medium. In a modest
way, this echoes Kelvin's hope that the vibrations of vortex rings would identify atoms through their
spectra,\cite{Kelvin1867} a hope that had faded by the time J.~J. Thomson, who had himself calculated these
vibrations, discovered the electron.

Part of the delay of the Barkley rings is simple to understand, because their drift speed grows more slowly
than their curvature. The remaining part surprised us. When we bent a ring out of its plane, its response
depended mainly on the wavelength of the deformation, so that rings of different sizes and shapes responded
alike if this wavelength was the same. Thin vortex filaments show a similar dependence, known since
Kelvin's work on their bending waves,\cite{Kelvin1880,Saffman1992} and there it arises from the
long-range flow that the vortex induces. For scroll rings, the origin remains open, and candidates are
the finite size of the spiral core and the waves that the filament emits.

Two approaches could settle this question. Computing the couplings from the response functions of the
spiral\cite{Henry2002,Dierckx2013,Biktasheva2009,Biktasheva2010} would show how much a local theory can
explain, and encasing a ring in a moving ``glass tube'', which removes long-range wave
interactions,\cite{Binysh2019} would test whether the emitted waves are responsible. Axis switching might
also be observable in the BZ reaction if the drift coefficient is not small compared with the tension, for
example near the transition to negative tension, using elliptic initiating wires and photomasks to prepare
the rings and optical tomography\cite{Winfree1996,Bansagi2008} to follow them. In the gelled reagent of Winfree and Jahnke, the estimated drift was less than half of the shrinkage,\cite{WinfreeJahnke1989}
so that a ring would shrink at least about threefold before its first switch. Finally, strongly elongated
vortex rings in fluid dynamics split,\cite{KambeTakao1971,Dhanak1981,Oshima1988,Cheng2016} and it will be interesting to see
whether highly eccentric scroll rings do the same.

\section*{Supplementary Material}
See the supplementary material for derivations of the results of Sec.~\ref{sec:theory},
including the rigidity extension; numerical controls, the filament definitions, and the filament
solver; the half-cycle analysis with all Barkley half-cycles and the
comparison of procedures; the circular-ring calibrations at $b = 0.15$ to 0.18 and the
two-dimensional spirals of all models; the shape-mode analysis with all coefficients and their
ranges (Tables~S5 and S6), fits, filament definitions, twist, the test of the rigidity terms, straight
filaments (Table~S10), the wavenumber dependence of $\beta_\zeta$ (Table~S11, Fig.~S3), and the comparison with
thin vortex rings; the definition and numerical treatment of the Panfilov--Rudenko model with its
circular and elliptic rings (Tables~S12--S13, Fig.~S4); the Atri model of Ca$^{2+}$ waves
(Tables~S14--S15, Fig.~S5); and our simulations of the Oregonator model.

\begin{acknowledgments}
This work was supported by the National Science Foundation under Grant No. CHE-2515818.
\end{acknowledgments}

\section*{Author declarations}
\subsection*{Conflict of Interest}
The author has no conflicts to disclose.

\subsection*{Author Contributions}
O.S.: Conceptualization; Funding acquisition; Investigation; Validation; Writing -- review \& editing.

\section*{Data availability}
The simulation and analysis code is openly available on GitHub at
\url{https://github.com/osteinbock/elliptic-scroll-rings}. The data that support the findings of this
study can be regenerated with this code.

\bibliography{refs}

\end{document}